\documentclass[12pt]{iopart}
\pdfoutput=1

\usepackage{iopams}

\usepackage[utf8]{inputenc}
\usepackage[T1]{fontenc}
\usepackage{lmodern}
\usepackage{dirtree}

\usepackage{amsfonts, amssymb, latexsym}
\usepackage{graphicx}
\usepackage{color}

\usepackage{breakurl}
\usepackage{url}

\usepackage[sort&compress,numbers]{natbib}

\usepackage[unicode,breaklinks]{hyperref}
\hypersetup{
    pdftitle={An introduction to the Einstein Toolkit},
    pdfkeywords={General Relativity, Numerical Relativity, Einstein Toolkit},
    colorlinks=false,            
    bookmarks=true,              
  }

\usepackage{listings}
\newcommand{\codename}[1]{\texttt{#1}}

\graphicspath{{figures/}}

\sloppypar
\begin{document}

\title{An Introduction to the Einstein Toolkit}

\author{Miguel Zilh\~ao$^1$, Frank Löffler$^2$}

\address{$^1$ Center for Computational Relativity and Gravitation, \\
              Rochester Institute of Technology, 74-2068, %
              Rochester, NY 14623, USA}

\address{$^2$ Center for Computation \& Technology, \\
              Louisiana State University, %
              Baton Rouge, LA, USA}

\eads{\mailto{mzilhao@astro.rit.edu}, \mailto{knarf@cct.lsu.edu}}

\begin{abstract}
We give an introduction to the Einstein Toolkit, a mature, open-source
computational infrastructure for numerical relativity based on the Cactus
Framework, for the target group of new users. This toolkit is composed of
several different modules, is developed by researchers from different
institutions throughout the world and is in active continuous development.
Documentation for the toolkit and its several modules is often scattered across
different locations, a difficulty new users may at times have to
struggle with. Scientific papers exist describing the toolkit and its methods
in detail, but they might be overwhelming at first.  With these lecture notes
we hope to provide an initial overview for new users. We cover how to obtain,
compile and run the toolkit, and give an overview of some of the tools and
modules provided with it.
\end{abstract}


\maketitle

\section{Introduction}
\label{sec:intro}

Dating back to Hahn and Lindquist's first attempts at numerically evolving
Einstein's field equations for a binary black hole spacetime~\cite{Hahn:1964},
numerical relativity is now an essential tool to study systems with strong and
dynamical gravitational fields.  Originally motivated essentially from
astrophysics and gravitational wave astronomy, it was soon realized that
numerical relativity could be useful to a much wider spectrum, with additional
motivations coming even from fields other than gravity itself. See for
instance~\cite{Cardoso:2012qm} for an overview of some of these topics.

Developing efficient computational codes to probe this highly non-linear regime
is a very non-trivial and time-consuming process, however.  In fact, only in
2005 were the first simulations of stable, long-term evolutions of the inspiral
and merger of two black
holes~\cite{Pretorius:2005gq,Baker:2005vv,Campanelli:2005dd} successfully
accomplished.
It is therefore a great advantage if well-tested, easy to use computational
infrastructures are available that allow researchers to focus less on the
computational aspect of such tasks, allowing for more time being spent on the
actual physics.

To illustrate the helpfulness of having such an infrastructure, we start by
noting that a typical numerical relativity code will have to, among others,
incorporate a mesh refinement algorithm, use an efficient parallelization
scheme, efficiently deal with large input and output and also use somewhat
complex tools for analysis.

In these notes we introduce the Einstein
Toolkit~\cite{EinsteinToolkit:web,Loffler:2011ay,Moesta:2013dna}, an open source,
community-driven, freely accessible infrastructure for numerical relativity.
The toolkit uses the BSSN (Baumgarte, Shapiro, Shibata, Nakamura) evolution
system for spacetime evolution~\cite{Baumgarte:1998te,Shibata:1995we} with a
finite-volume general-relativistic hydrodynamics solver.
Within this document, however, we have chosen
to only describe components and tools that we believe are of most interest
for a new user, necessarily leaving out large parts of the toolkit that are
of interest for advanced users. For a more detailed description the interested
reader is directed to~\cite{Loffler:2011ay}, and for an overview of the
MHD capabilities to~\cite{Moesta:2013dna}.

This document is structured as follows. We start with a high-level overview of
the most important modules for new users in section~\ref{sec:overview}, before
we describe basic steps to use the Einstein Toolkit in section~\ref{sec:usage}.
Some details about the structure of most modules are given in
section~\ref{sec:thorn_anatomy}, which is not only interesting for users to
understand how to configure simulations, but also for developers that are
interested in extending the toolkit. Finally, some examples show-case Einstein
Toolkit usage in section~\ref{sec:examples}, having in mind that these
examples can be used within tutorials or workshops. In particular, this means
that these are not tailored for the best possible physical results, but instead
for short run-time and low system requirements. Finally, we conclude in
section~\ref{sec:remarks}.

\section{Structural Overview}
\label{sec:overview}

The Einstein Toolkit (ET) consists of quite a large number of components, of which
most currently use the \codename{Cactus} Computational
Toolkit~\cite{Cactuscode:web,Goodale:2002a} framework, which provides the basic
modular infrastructure for numerical simulations. A number of tools are
provided surrounding this framework environment, from simulation management to
data analysis and visualization. In the following we briefly introduce
{\ttfamily Cactus}, as well as some of the arrangements provided with the
toolkit. Some of the more important tools are introduced later in
section~\ref{sec:usage}, as part of a description of typical Einstein Toolkit
usage.

\subsection{Cactus}
\label{sec:cactus}

The \codename{Cactus}
framework~\cite{CactusUsersGuide:web,Goodale:2002a,Cactuscode:web} is a general
framework for the development of portable, modular applications, wherein
programs are split into components (called \emph{thorns}) with clearly defined
dependencies and interactions.  Thorns are typically developed independently
and should be interchangeable with others with same functionality. They usually do not
directly interact with each other; rather each of them interacts with the
\codename{Cactus} framework \emph{flesh} which provides the ``glue'' between different
thorns.  Thorns can be written in different languages, with \codename{C},
\codename{C++}, \codename{Fortran}, \codename{OpenCL} and \codename{CUDA} being supported currently.

\codename{Cactus}, originally developed for numerical relativity, is a direct descendant
of many years of code development in Ed Seidel's group of researchers at NCSA;
its version 1.0 was released in 1995.  Over the years it has been generalized
for use by scientists in other domains.

At this point, introducing some \codename{Cactus}-related terminology is
beneficial, making text within this publication compatible to existing
documentation within the Einstein Toolkit.

\codename{Cactus} simulations require an executable to be compiled, which is
done typically by the users themselves. This setup allows for quick adaptation
to changes in the local environment, e.g., updated libraries, but also for a
simple way to tailor the source code to the users' needs, most of the time
through usage of own modules. This executable has one mandatory argument: a
\emph{parameter file}. This is a simple text file, containing key-value pairs
of parameter names and the desired settings within the simulation. An
executable from all thorns within the Einstein Toolkit can be used to model a
variety of physical scenarios. The parameter file is used to choose which of
these should be realized.

An often used term within \codename{Cactus} is a so-called \emph{grid function}. This is a (discretization of a)
variable which is defined on every point of a given grid. Examples are the
rest-mass density of a fluid which is defined in all cells within a given
domain, or the components of the metric tensor in general relativity. In
parallel simulations, a grid function is typically split across processes,
allowing storage of large grid functions and parallel work on each part.

\subsection{Carpet}
\label{sec:carpet}

\texttt{Cactus} separates physics code from infrastructure code, i.e., a
typical physics thorn will not contain any memory management, parallelization,
Input/Output (IO) or mesh refinement code. Most of these tasks are bundled in one
special thorn, the so-called \emph{driver thorn}, or driver.

Two drivers are provided with the ET, \codename{PUGH} and \codename{Carpet}.
\codename{PUGH} implements a uniform Cartesian grid, while
\codename{Carpet}~\cite{CarpetCode:web,Schnetter:2003rb, Schnetter:2006pg} provides
more than unigrid: Berger-Oliger style~\cite{Berger:1984zza} adaptive
mesh refinement (AMR). Its capabilities include
\begin{itemize}
  \item splitting grid functions and arrays among the MPI processes,
  \item setting up a mesh refinement grid hierarchy,
  \item communicating grid function information between MPI processes,
  \item communicating between refinement levels by prolongation and
        restriction,
  \item modifying grid hierarchy (regridding) when requested,
  \item performing parallel IO.
\end{itemize}
Together with the \codename{Llama} code~\cite{Pollney:2009yz}, Carpet can
provide multiblock infrastructure, providing different patch systems that cover
the simulation domain by a set of overlapping patches. \codename{Llama} is
publicly available, but, as of writing, not yet part of the
Einstein Toolkit.

For the sake of simplicity, and because of the almost exclusive usage of
\codename{Carpet} within numerical relativity using the Einstein Toolkit, we
will in the following concentrate exclusively on examples using the AMR mesh
refinement driver, without multiple patches.

\subsection{Arrangements}

\emph{Arrangements} are collections of \emph{thorns}, usually signaling a
common task, or a common origin. This grouping is solely done for the
benefit of a better overview for the user. Being part of one or another
arrangement does not have any special meaning for a \codename{Cactus} thorn
beyond that.

Let us first give an overview on some of the basic \codename{Cactus} and ET arrangements,
before we present some details on a few of the more important thorns.

\subsubsection{Core Cactus arrangements}
The main core \codename{Cactus} arrangements include the following.
\begin{description}
\item[\texttt{CactusBase}] \hfill \\
  Provides infrastructure thorns for boundary conditions, setting up the coordinates, Input and Output, symmetries and time.
\item[\texttt{CactusNumerical}] \hfill \\
  Provides numerical infrastructure thorns for time integration, artificial
  dissipation, symmetry boundary conditions, setting up spherical surfaces,
  interpolation, Method of Lines (MoL) implementation (see section~\ref{sec:mol}
  for more details), and many others.
\item[\texttt{CactusUtils}] \hfill \\
  Provides some utility thorns: Formaline (see section~\ref{sec:formaline} for
  details), nan-checking, termination triggering and timer reports.
\item[\texttt{ExternalLibraries}] \hfill \\
  Provides external libraries---LAPACK, GSL, HDF5, FFTW, Lorene (for initial
  data), MPI, among others---that are automatically configured and compiled if they
  are not found on the user's machine.
\end{description}

\subsubsection{EinsteinBase}
\label{sec:einsteinbase}
  
\begin{figure}
 \centering
 \includegraphics[width=0.5\textwidth]{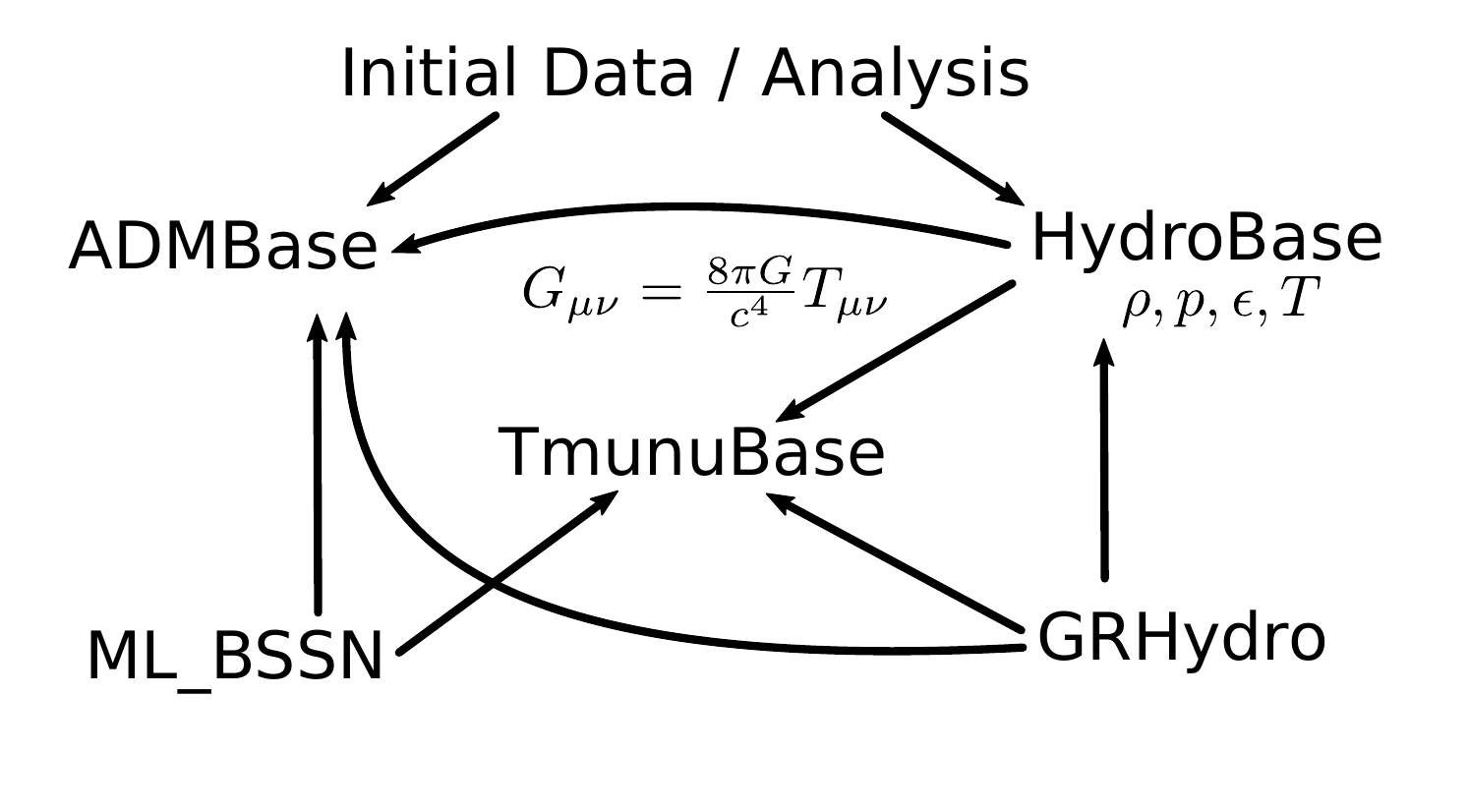}
 \caption{Simplified overview over the central thorns of \texttt{EinsteinBase}:
 \texttt{ADMBase}, \texttt{HydroBase}, and \texttt{TMunuBase}, together with
 some example ``user'' thorns, and their dependencies. In this case ``user'' thorns
 are \texttt{ML\_BSSN} for curvature evolution,
 \texttt{GRHydro} for magneto-hydrodynamics evolution and initial data and analysis
 thorns, all not directly depending on the evolution thorns, nor on each other.
 \label{fig:EinsteinBase}}
\end{figure}

\texttt{EinsteinBase} thorns define and register basic variables within
numerical relativity. Thorns that make use of or modify, for example, the
ADM~\cite{Arnowitt:1962hi,York:1979sg}
variables, or the stress-energy tensor, should inherit these thorns instead of defining their own. See Figure~\ref{fig:EinsteinBase} for a sketch showcasing typical relations between these thorns and other ``user'' thorns.
Almost all thorns relating to the physics of
numerical relativity inherit from some, if not all of these thorns. It
is vital even to a new user to have at least an overview of their structure.
Main thorns include:

\begin{description}

\item[\texttt{ADMBase}] \hfill \\
Defines groups of grid functions for basic spacetime variables, based
on the $3+1$ ADM construction~\cite{Arnowitt:1962hi,York:1979sg},
which makes it the natural choice of a common foundation for
exchanging data between modules using different formalisms.
Concretely, \texttt{ADMBase} defines:
\begin{itemize}
 \item 3-metric $\gamma_{ij}$ ({\tt gxx, gxy, gxz, gyy, gyz, gzz}),
 \item extrinsic curvature $K_{ij}$ ({\tt kxx, kxy, kxz, kyy, kyz, kzz}),
 \item lapse $\alpha$ ({\tt alp}), shift $\beta^{i}$ ({\tt betax, betay, betaz}),
 \item time derivative of lapse $\partial_t \alpha$ ({\tt dtalp}) and time derivative of
       shift $\partial_t \beta^i$ ({\tt dtbetax, dtbetay, dtbetaz}).
\end{itemize}
\texttt{ADMBase} also defines basic parameters to choose the initial data,
evolution method and other details common to a number of other thorns, e.g.,
schedule groups (specific points in time within a simulation) for other thorns
to schedule their routines modifying the \texttt{ADMBase} variables.
We stress that, by inheriting these variables from \texttt{ADMBase}, different thorns (typically written by different people) are then able to interact smoothly.

\item[\texttt{HydroBase}] \hfill \\
Defines basic variables and grid functions for hydrodynamics evolutions, e.g.:
\begin{itemize}
  \item rest mass density $\rho$ ({\tt rho}),
  \item pressure $P$ ({\tt press}),
  \item specific internal energy $\epsilon$ ({\tt eps}),
  \item contravariant fluid three velocity $v^i$ ({\tt vel[3]}),
  \item Lorentz factor $W$ ({\tt w\_lorentz}),
  \item electron fraction $Y_e$ ({\tt Y\_e}),
  \item temperature $T$ ({\tt temperature}),
  \item specific entropy per particle $s$ ({\tt entropy}),
  \item contravariant magnetic field vector $B^i$ ({\tt Bvec[3]}).
\end{itemize}
\texttt{HydroBase} also sets up scheduling groups for recommended interaction
with the fluid variables.

\item[\texttt{TmunuBase}] \hfill \\
Defines grid functions for stress-energy tensor (``right-hand-side'' of Einstein's equations):
\begin{itemize}
  \item time component $T_{00}$ ({\tt eTtt}),
  \item mixed components $T_{0i}$ ({\tt eTtx}, {\tt eTty}, {\tt eTtz}),
  \item spatial components $T_{ij}$ ({\tt eTxx}, {\tt eTxy}, {\tt eTxz},
                                         {\tt eTyy}, {\tt eTyz}, {\tt eTzz}).
\end{itemize}
\texttt{TmunuBase} also sets up scheduling groups for other thorns to schedule
routines that adds to the stress-energy tensor. 

\end{description}

\subsubsection{EinsteinInitial}

Arrangement \texttt{EinsteinInitial} contains several initial data thorns. The most widely used of these include the following:
\begin{description}
  \item[\texttt{TwoPunctures}] \hfill \\
    Very efficient pseudo-spectral code that computes puncture-type binary black hole initial data~\cite{Ansorg:2004ds}.
  \item[\texttt{TOVSolver}]  \hfill \\ 
    Initial data for a single TOV star.
\end{description}

\subsubsection{EinsteinEvolve / McLachlan}
These arrangements contain several thorns related with the numerical evolution of Einstein's field equations:
\begin{description}
\item[\texttt{McLachlan}]  \hfill \\
  \texttt{McLachlan}~\cite{Brown:2008sb,Reisswig:2010cd,McLachlan:web} is ET's
  spacetime evolution code. It uses a accurate finite differencing scheme up to
  eighth order with corresponding Kreiss-Oliger dissipation terms, to discretize
  spacetime variables in the BSSN
  form\cite{Shibata:1995we,Baumgarte:1998te,Alcubierre:2000xu}. It is designed
  to inter-operate through the \texttt{ADMBase} and \texttt{TmunuBase}
  interface. All \texttt{McLachlan} thorns are auto-generated from tensor
  equations via \texttt{Kranc}~\cite{Husa:2004ip,Lechner:2004cs,Kranc:web}, a
  \texttt{Mathematica} application which converts a high-level continuum
  description of a PDE into a highly optimized module for {\tt
    Cactus}.\footnote{We note that \texttt{Mathematica} is not needed to compile
    the thorn.}

\item[\texttt{GRHydro}]  \hfill \\
  General relativistic magneto-hydrodynamics (GRMHD) matter
  evolution~\cite{Baiotti:2004wn,Hawke:2005zw,Moesta:2013dna}, also designed to inter-operate
  through the \texttt{ADMBase} and \texttt{TmunuBase} interface.

\item[\texttt{PunctureTracker}] \hfill \\
  Takes care of black hole tracking, e.g., to allow the automatic steering of
  mesh refinement grids according to black hole positions.

\item[\texttt{NewRad}] \hfill \\
  Implements radiative, Sommerfeld-type, boundary conditions on specified grid functions~\cite{Alcubierre:2002kk}.
\end{description}

\subsubsection{EinsteinAnalysis}

This arrangement includes several thorns useful for analysis:
\begin{description}
\item[\texttt{AHFinderDirect}]  \hfill \\
  This thorn finds black hole apparent horizons\footnote{More generically,
    \texttt{AHFinderDirect} looks for a closed 2-surface with $S^2$ topology
    having any desired constant expansion.} given the 3-metric and extrinsic
  curvature~\cite{Thornburg:2003sf,Thornburg:1995cp}. \texttt{AHFinderDirect} is very fast and accurate, but requires an
  initial guess for the apparent horizon position.
\item[\texttt{WeylScal4}]  \hfill \\
  Calculates the Weyl scalar $\Psi_4$, often used within the context of the
  Newman-Penrose formalism~\cite{Newman:1961qr}. $\Psi_4$, in the appropriate
  frame, can be shown to encode the outgoing gravitational radiation of an
  asymptotically flat system, making this quantity a very useful one for
  gravitational wave analysis. \texttt{WeylScal4} is auto-generated from
  tensor equations via \texttt{Kranc}.
\item[\texttt{Multipole}]  \hfill \\
  Decomposes arbitrary grid functions into spin-weighted spherical
  harmonics. Often used in combination with \texttt{WeylScal4}.
\item[\texttt{ADMAnalysis}]  \hfill \\
  Calculates several quantities from the ADM variables. In particular it can
  compute, if requested,
  \begin{itemize}
  \item the trace of the extrinsic curvature ({\tt trK});
  \item the determinant of the metric ({\tt detg});
  \item the components of the metric and extrinsic curvature in spherical
    coordinates ({\tt grr}, {\tt grq}, {\tt grp}, {\tt gqq}, {\tt gqp}, {\tt
      gpp}, {\tt Krr}, {\tt Krq}, {\tt Krp}, {\tt Kqq}, {\tt Kqp}, {\tt Kpp});
  \item the components of the Ricci tensor ({\tt Ricci11}, {\tt Ricci12}, {\tt
      Ricci13}, {\tt Ricci22}, {\tt Ricci23}, {\tt Ricci33}) and the Ricci
    scalar ({\tt Ricci}).
  \end{itemize}
\item[\texttt{EHFinder}]  \hfill \\
  Finds event horizons in numerical spacetimes by integrating a null surface
  backwards in time. \texttt{EHFinder} is a post-processing analysis thorn. It
  is therefore necessary to evolve the initial data forward in time, beforehand,
  while outputting enough 3D data. The 3D data is then read back in, in reverse
  order. {\tt EHFinder} can follow the event horizon during the merger of two
  (or more) black holes into one black hole. For more information, we refer
  to~\cite{Diener:2003jc} and to the thorn's documentation.
\item[\texttt{ADMConstraints / ML\_ADMConstraints}]  \hfill \\
  \texttt{ADMConstraints} calculates the ADM constraints violation
  (\texttt{ham}, \texttt{hamnormalized},
  \texttt{momx}, \texttt{momy}, \texttt{momz}) from the ADM variables. Note
  that \codename{McLachlan} contains thorn \codename{ML\_ADMConstraints} which
  calculates the same quantities, but directly using potentially
  higher-order derivatives, and is, in general, preferred over ADMConstraints.
\end{description}

\subsection{Tools}

Having briefly described some of the main arrangements provided with ET, let us
now turn our attention to some tools that are provided as well.

\subsubsection{SimFactory}

\emph{SimFactory} (for Simulation Factory), to be discussed in more detail in section~\ref{sec:simfactory}, is a tool that incorporates a set of abstractions for tasks which are necessary to successfully use the \texttt{Cactus} framework. These abstraction layers include:
\begin{itemize}
\item accessing remote systems and synchronizing source code trees;
\item configuring and building on different systems semi-automatically;
\item providing maintained list of supercomputer configurations;
\item managing simulations (follow ``best practices'' and avoid human errors).
\end{itemize}

\subsubsection{GetComponents}

As we have seen, the ET is composed of several different independent components (\texttt{Cactus}, \texttt{Carpet}, and several thorns).
Typically, these different parts are hosted at different repositories, with different version control systems, and are maintained by different groups.
\texttt{GetComponents}~\cite{Seidel:2010aa,Seidel:2010bb} provides a unified way of downloading all of these components in a user-friendly way, hiding this complexity from the user.

\texttt{GetComponents} is a script with one mandatory argument, a so-called
thornlist. This is a simple text file describing the locations and methods
of how to retrieve components. The ET provides one such thornlist, describing
these details for all its components (see section~\ref{sec:download_and_running}).
This thornlist can be either
given as URL, or as local file name. The most interesting optional parameter to
\texttt{GetComponents} for new users is probably \verb|--parallel|, to enable
parallel checkouts, reducing the time to retrieve the whole toolkit considerably
at times. Help and a list of other options can be obtained via the \verb|--help|
option.

\subsubsection{Formaline}
\label{sec:formaline}

Results obtained via computation have to be repeatable. However, ensuring this
in practice can be a tedious task. One tool which can greatly help here is
\texttt{Formaline}. All a user has to do is to compile and active this thorn.
\texttt{Formaline} then:
\begin{itemize}
\item Ensures that simulations are and remain repeatable, remembering
  exactly how they were performed.
\item Takes snapshots of source code, system configuration; stores it in
  executable and/or git repository.
\item Tags all output files.
\end{itemize}

It should be noted that the usage of \texttt{Formaline} is in no way a
guarantee that simulations are, in fact, repeatable exactly. Almost no
simulation can be repeated bit-identical due to the parallel nature of some
algorithms and their inherent ``random'' numerical errors. However,
\texttt{Formaline} automates tasks a user \textit{can} do in such a convenient
way that most users actually \textit{do} them.

\section{Using the ET}
\label{sec:usage}

\subsection{Requirements}

\texttt{Cactus} and the Einstein Toolkit should run on all major variants of
Unix. Main requirements are the following: \texttt{C}, \texttt{C++} and
\texttt{Fortran 90} compilers; an MPI implementation (e.g., OpenMPI, MPICH; will
be provided by the ET if not found); HDF5 (will also be provided if not found);
\texttt{Perl} and \texttt{Python}. Obtaining the ET directly from its
repositories also requires \texttt{Subversion} and \texttt{git} tools. All of
these requirements are usually conveniently available via the respective
distribution repositories, or are commonly available on supercomputers.

In the following we will be assuming that all such tools are available.

\subsection{Downloading and Running}
\label{sec:download_and_running}

This tutorial will be based on the latest stable ET release, which at the time
of writing is ``\O{}rsted'' (released on November 8th, 2012). We expect a new
release of the ET to be published between submission, and publication of this
article. Exchanging the string \verb|ET_2012_11| by \verb|ET_2013_05| in the
following should be all that is needed to use the new release.

To download the ET, one must first obtain the \verb|GetComponents| script and point it to the appropriate \emph{thornlist}:
\begin{lstlisting}
curl -O https://raw.github.com/gridaphobe/CRL/master/GetComponents 
perl GetComponents https://svn.einsteintoolkit.org/manifest/branches/ET_2012_11/einsteintoolkit.th
\end{lstlisting}
A \emph{thornlist} is a text file containing the list of thorns to be downloaded together with the path and version controlling system used for each thorn.
Such a file will typically have the following structure:
\begin{lstlisting}
!CRL_VERSION = 1.0

!DEFINE ROOT = Cactus
!DEFINE ARR  = $ROOT/arrangements
!DEFINE COMPONENTLIST_TARGET = $ROOT/thornlists/

!DEFINE ET_RELEASE = ET_2012_11

# Cactus Flesh
!TARGET   = $ROOT
!TYPE     = svn
!AUTH_URL = https://svn.cactuscode.org/flesh/branches/$ET_RELEASE
!URL      = http://svn.cactuscode.org/flesh/branches/$ET_RELEASE
!CHECKOUT = Cactus
!NAME     = .

# Cactus thorns
!TARGET   = $ARR
!TYPE     = svn
!AUTH_URL = https://svn.cactuscode.org/arrangements/$1/$2/branches/$ET_RELEASE
!URL      = http://svn.cactuscode.org/arrangements/$1/$2/branches/$ET_RELEASE
!CHECKOUT =

CactusBase/Boundary
CactusBase/CartGrid3D
CactusBase/CoordBase
CactusBase/Fortran
CactusBase/InitBase
CactusBase/IOBasic
CactusBase/IOUtil
CactusBase/Time

(...)
\end{lstlisting}
The user is also free to download the list first to a file and then add its own
private thorns.
The \verb|GetComponents| script accepts a thornlist file as an argument. To download all the thorns specified in the \verb|einsteintoolkit.th| file (in this example using the \verb|--parallel| option to retrieve repositories in parallel to speed-up the process):
\begin{lstlisting}
perl GetComponents --parallel einsteintoolkit.th
\end{lstlisting}
Once \verb|GetComponents| finishes downloading one should have a folder called \verb|Cactus| with the following structure:
{\small\dirtree{%
.1 Cactus/.
.2 arrangements/.
.2 bin/.
.2 doc/.
.2 lib/.
.2 manifest/.
.2 par/.
.2 repos/.
.2 simfactory/.
.2 src/.
.2 thornlists/.
.2 utils/.
.2 CONTRIBUTORS.
.2 COPYRIGHT.
.2 Makefile.
}}
We are now ready for the configuring and compiling stage.  The first step here
is choosing the config (\verb|*.cfg|) file for the machine. A few examples (which
include the configuration files for Fedora, Debian, Ubuntu and OS~X) are
provided under \verb|./Cactus/simfactory/mdb/optionlists/|.  The second step is
having the required thornlist for the configuration intended to be built.
Different configurations, compiled with different thornlists, are free to
co-exist. Note that, typically, one will \emph{not} be compiling \emph{all} the
thorns provided with the ET.  Compilation is time-consuming, and different
configurations also take a significant amount of disk space.  One therefore
typically builds a thornlist that is as small as possible, including only the
required thorns. Care should be taken, though, as there are often non-trivial
dependencies between thorns.  If one thorn which is required by another thorn
is not mentioned in the thornlist, compilation will abort (with
the corresponding error message).

The traditional way of compiling \texttt{Cactus} (i.e., without
SimFactory---see below) is as follows. First, the configuration is created
\begin{lstlisting}
cd Cactus
make ET-config options=<machine config file> THORNLIST=<thornlist>
\end{lstlisting}
This creates a configuration called ``ET'', but any other name could be chosen
here. Once the configuration is done, the compilation process is simply
\begin{lstlisting}
make -j <number of processes> ET 
\end{lstlisting}
If everything is compiled correctly, the executable \verb|cactus_ET| will be
created under \verb|./exe/|.  These steps need to be repeated for every
different configuration (typically, with different thornlists) built.

We should now be ready for running. For this, a parameter file is needed, specifying which thorns to use within the simulation (not all compiled thorns need to be active), and which specific model parameters have been chosen. A few examples are provided under \verb|./Cactus/par/|, including the parameter file for an inspiraling collision of black holes in vacuum (\verb|qc0-mclachlan.par|) as well as the parameter file for a static TOV star (\verb|static_tov.par|).
The typical procedure for running, as with other MPI executables, is
\begin{lstlisting}
mpirun -np <num procs> ./exe/cactus_ET <parameter file>
\end{lstlisting}

\subsection{Analyzing the output}

Output is typically controlled by the following thorns:
\begin{description}
\item[\texttt{CarpetIOASCII}] \hfill \\
  writes 0, 1, 2 or 3D output from the specified variables onto a text file (\verb|.asc|)
\item[\texttt{CarpetIOScalar}] \hfill \\
  performs scalar reductions (maximum, minimum, norm, \dots) of the specified variables and writes onto a text file (\verb|.asc|);
\item[\texttt{CarpetIOHDF5}]  \hfill \\
  writes 0, 1, 2 or 3D output from the specified grid functions onto an HDF5 file; also handles check pointing (\verb|.h5|).
\end{description}
All of these thorns include, among others, the following parameters: \verb|out?D_vars|, \verb|every?D_vars|, where \verb|?=0,1,2,3| for 0-, 1-, 2- or 3-dimensional output respectively.
\verb|out?D_vars| is a list of grid functions to output at every \verb|every?D_vars| iteration.
3-dimensional output is by default produced over the entire grid; 2-dimensional output is given at specified 2D-planes (the default being the \texttt{xy}, \texttt{xz} and \texttt{yz} planes containing the origin); 1-dimensional output is given along lines (default being the \texttt{x}, \texttt{y} and \texttt{z} axis); 0-dimensional output is given at specified points (default being the coordinate origin).

For visualizing 1-dimensional ASCII output, standard tools like \texttt{matplotlib}, \texttt{gnuplot} and \texttt{xmgrace} are often used; for 2- and 3-dimensional HDF5 output, \texttt{VisIt} and \texttt{DV} are popular (freely available) options.


\section{Anatomy of a Cactus thorn}
\label{sec:thorn_anatomy}

The knowledge about the structure of \texttt{Cactus} thorns is strictly speaking
not necessary for new users, as information about their interface can be
obtained via built documentation.
However, a short overview is given here to illustrate how first steps towards
development within the toolkit look like.

Any given \texttt{Cactus} thorn will have the following directory structure (under \verb|./Cactus/arrangements/|)
{\small\dirtree{%
.1 ArrangementName/ThornName/.
.2 COPYRIGHT.
.2 README.
.2 test/.
.2 doc/.
.2 par/.
.2 configuration.ccl.
.2 interface.ccl.
.2 schedule.ccl.
.2 param.ccl.
.2 src/.
.3 make.code.defn.
}}
\verb|COPYRIGHT|, \verb|README|, \verb|doc/|, \verb|test/| and \verb|par/| are optional; \verb|doc/| will be typically used to store some documentation related with the thorn, and \verb|par/| can have example parameter files for the thorn.
Under \verb|src/|, the file called \verb|make.code.defn| is needed with the list of files to be compiled.

\subsection{Cactus configuration files}

Knowledge of these files is strictly only necessary for developers, as information within is automatically included in build documentation. However, we include a short description of their contents here because even for a new user it can be useful at times to have at least a general overview of the interface of \codename{Cactus} thorns with the flesh.

\codename{Cactus} thorns use four files to specify their interface with the \codename{Cactus} flesh (three compulsory and one optional). They can be found in the top level thorn directory, using the \texttt{*.ccl} file name extension (\codename{Cactus} Configuration Language,
see~\cite{Seidel:2010bb} for extensive details).

The file \verb|configuration.ccl| (optional) lists inter-thorn build dependencies.
\verb|interface.ccl| is used to define thorn-wide variables, grid functions, and shared functions. 
It consists of a header block detailing the thorn's relationship with other thorns, 
a block stating which include files are used from other thorns, and which include files are provided by this thorn, blocks with aliased functions provided or used by this thorn and a series of blocks listing the thorn's global variables.
Note that functions can be called by a different name within a given thorn.
\verb|schedule.ccl| takes care of all the function scheduling and controls the global storage of grid functions.
And finally, \verb|param.ccl| defines all parameters (which, in contrast to regular variables can be set at start-time) and sets their default values.

\subsection{Thorn Examples}
In the following, we present a few examples of thorns within the Einstein Toolkit, with the aim of deepening the understanding of \codename{Cactus} thorns obtained in previous sections.

\subsubsection{ADMBase}

This thorn, already briefly covered in section~\ref{sec:einsteinbase}, is
fundamental for most (if not all) numerical relativity simulations, providing
the core infrastructure for thorns implementing general relativity on a 3D grid
in the 3+1 formalism. \codename{ADMBase} provides the basic variables (3-metric,
extrinsic curvature, lapse function and shift vector) for the 3+1 formalism.
These grid functions are then inherited by every thorn using them. This allows
a well-defined interaction between thorns implementing fundamentally different
functionality, e.g., thorns providing initial data, evolution methods and
analysis routines for the 3+1 formalism. 

The ET uses this thorn, ensuring that different analysis and initial data
thorns are able to communicate with each other and evolution thorns,
independently of the variables used for the evolution (typically, \emph{not}
the ADM variables).  Generically, an initial data thorn will initialize the
\codename{ADMBase} variables to the initial data describing the desired
physical configuration.  The evolution thorn will then import the
\codename{ADMBase} variables, convert them into its evolution variables (the
BSSN variables, for example) and perform the evolution using these variables.
At every time step, the \codename{ADMBase} variables are reconstructed from the
evolved variables so that they always contain the current solution.  Analysis
thorns such as those for horizon-finding or wave extraction then use the
\codename{ADMBase} variables as well. This effectively decouples initial-data
solvers, evolution methods and analysis routines, making them inter-changeable
and compatible with each other.

\subsubsection{HydroBase}

Similarly to \codename{ADMBase}, \codename{HydroBase} defines and stores the
primitive variables common among hydrodynamic simulations, needed parameters
and schedule groups for the main functions of a hydrodynamics code.
\codename{HydroBase} does not contain the actual source code of typical
routines of hydrodynamics codes; it merely provides a common setup in which
hydrodynamics codes can schedule their routines.

This way, different modules of hydrodynamics codes (such as initial data solvers
or analysis modules) working only with entities defined in \codename{HydroBase}
can be used interchangeably. Yet another advantage is that the output generated
by different hydrodynamics codes within \codename{Cactus} would be the same,
including variable names and unit conventions, thus improving the ability to
compare results from different codes.

\codename{HydroBase} also sets up scheduling blocks organizing main functions
which modules of a hydrodynamics code may need. These scheduling blocks are
optional, but when used may simplify existing codes and make them more
inter-operable.
Currently the scheduling blocks are:
\begin{itemize}
 \item initializing the primitive variables;
 \item converting primitive variables to conservative variables;
 \item calculating the right hand side in the method of lines (MoL);
 \item setting and updating an excision mask;
 \item applying boundary conditions.
\end{itemize}
This way, initializing the primitive variables, recovering the conservative
variables, and basic atmosphere handling can be implemented in different thorns
while allowing a central access point for analysis thorns.

\subsubsection{SphericalSurface}

\codename{SphericalSurface} defines two-dimensional surfaces with spherical
topology.  The thorn itself only acts as a repository for other thorns to set
and retrieve such surfaces, making it a pure infrastructure thorn. Different
thorns can then update a given spherical surface, while others then read
that information without having to know about the first thorn.

Within the ET, uses of spherical surfaces include the following: storing
apparent horizon information (used by \codename{AHFinderDirect}); tracking
black hole location (by thorn \codename{CarpetTracker}), information which is
then used to determine where to perform mesh refinement.

The thorn provides, for each surface, a two-dimensional grid array
\texttt{sf\_radius} and grid scalars \texttt{sf\_origin\_x},
\texttt{sf\_origin\_y}, and \texttt{sf\_origin\_z}.  The number of surfaces is
determined by the parameter \texttt{nsurfaces}, which has to be set in the
parameter file.

\subsubsection{Multipole}
\label{sec:multipole}

\codename{Multipole} is a thorn that decomposes any \codename{Cactus} grid
function into spherical harmonics on coordinate spheres of given radii. One
application is to use it in combination with the \codename{WeylScal4} analysis
thorn to produce the mode decomposition of the Weyl scalars, often used in
numerical relativity.

Specifically, a grid function $u(t, r, \theta, \varphi)$ is expanded in
spin-weight $s$ spherical harmonics:
\begin{equation*}
  u(t, r, \theta, \varphi) = \sum_{l=0}^\infty \sum_{m=-l}^l C^{lm}(t,r) {}_s Y_{lm}(\theta,\varphi),
\end{equation*}
where the coefficients $C^{lm}(t,r)$ are given by
\begin{equation*}
C^{lm}(t, r) = \int {}_s Y_{lm}^* u(t, r, \theta, \varphi) r^2 d \Omega .
\end{equation*}
The thorn computes $C^{lm}(t, r)$ for the requested grid functions at given coordinate radii $r_i$.

To use this thorn, one specifies the grid functions to decompose, as well as the number and radii of the extraction spheres, as in the following example parameter file:
\begin{lstlisting}
ActiveThorns = "AEILocalInterp Multipole"
Multipole::nradii       =  3
Multipole::radius[0]    = 10
Multipole::radius[1]    = 20
Multipole::radius[2]    = 30
Multipole::variables    = "MyThorn::u"
\end{lstlisting}
The default parameters will compute all $l = 2$ modes assuming a
spin-weight $s = 0$ on every iteration of the simulation.\footnote{Note that for gravitational wave extraction, one will typically be interested in a spin-weight $s = -2$ decomposition.}  The
coefficients $C^{lm}$ will be output in files with names of the
form \verb|mp_<var>_l<lmode>_m<mmode>_r<rad>.asc|:
\begin{lstlisting}
mp_u_l2_m2_r10.00.asc
mp_u_l2_m-1_r20.00.asc
\end{lstlisting}

Note that, unlike most thorns, and since the spherical harmonic coefficients
are defined on coordinate spheres (and \emph{not} on the whole grid),
\codename{Multipole} takes care of the output by itself.  There is thus no need
to give \codename{Multipole} grid functions to \codename{CarpetIOASCII} or
\texttt{CarpetIOHDF5} in order to have output.

\subsubsection{MoL}
\label{sec:mol}

The Method of Lines (MoL) is a technique for numerically integrating time
dependent partial differential equations (PDEs).  The method discretizes only
the spatial derivatives of the PDE, thereby converting it into a large system
of ordinary differential equations (ODEs). This system can then be integrated
using well-known methods (such as Runge-Kutta and Crank-Nicolson).  Consider
the following PDEs system

\begin{equation*}
  \partial_t {\bf q} + {\bf A}^i({\bf q}) \partial_i {\bf B}({\bf q})
  = {\bf s}({\bf q}) ,
\end{equation*}
which we re-write in the form
\begin{equation}
  \label{eq:MoL}
  \partial_t {\bf q} = {\bf L}({\bf q}) .
\end{equation}
Assuming that the right-hand side is discretized, this is nothing but a system
of ODEs, which, as we mentioned, can be integrated using known ODEs
integrators.

The \codename{MoL} thorn provides a simple interface that implements this
technique inside \codename{Cactus}. We will here provide only some basic
information about the thorn; for more information we refer to its
documentation.

The idea behind the \codename{MoL} thorn is that the user should not be
concerned with writing the integration procedure itself.  Instead, in the
user's evolution thorn, grid functions that are to be evolved are
\emph{flagged} as \emph{evolved} variables, and their right-hand side (cf.
equation~(\ref{eq:MoL})) is given.  The \codename{MoL} thorn is then called where
appropriate and takes care of the evolution procedure itself, using the method
chosen by the user (currently supported include Runge-Kutta of different
orders, Iterative Crank-Nicolson (with and without averaging)).

For \codename{MoL}, grid functions need to be split into four categories:
variables for which we have a time evolution equation are \emph{evolved};
variables which the thorn sets but does not evolve are \emph{constrained}; any
other variable which the thorn reads during evolution is a \emph{save and
restore}  variable; other variables can be ignored.  Bear in mind that,
generically, one can have different variables being evolved by different
thorns. A grid function can thus be an \emph{evolved} variable from the point
of view of one thorn and a \emph{save and restore} from the point of view of
another---each thorn should register the grid function as they see it.  An
example would be the metric components in a general relativistic hydrodynamics
evolution: the hydrodynamics thorn will see the metric as a \emph{save and
restore} variable whereas the spacetime evolution thorn will see the metric as
\emph{evolved.} The \codename{MoL} thorn takes care automatically of the right
treatment of each variable.

In section~\ref{sec:wavemol} we will present a simple example of how to use
\codename{MoL}, to see how this procedure works in practice when evolving the
wave equation.

\subsection{Using SimFactory}
\label{sec:simfactory}

The \emph{Simulation Factory}~\cite{Thomas:2010aa,SimFactory:web} is designed to
manage several tasks necessary to set up and execute successful numerical
simulations using \codename{Cactus}.  Consider the following, very typical,
scenario: an ET user performs all his code development locally on a laptop,
where a few simple low resolution runs can be performed. To test more time
consuming cases the user may use a local workstation, which implies syncing the
code and parameter files, compiling and running the code on the workstation, and
then analyzing the output.  For higher resolution cases, the user may have to
resort to computer clusters, and the whole process of syncing, compiling,
running and analyzing will be repeated.  It is easy to see that this process can
be very error prone.

SimFactory provides an easy abstraction layer to this process, allowing the
user to do all his code development locally, and then syncing and compiling to
all other machines he may use with simple commands.  A record is also kept of
where and what simulations have been run.

To use SimFactory, we must first configure it. For machines known to SimFactory
this step consists essentially in configuring the user name and email address
(for job submission status emails). If a given machine configuration is not
already stored, SimFactory also needs an optionlist to use for the ET build.

\begin{lstlisting}
cd Cactus
./simfactory/bin/sim setup --optionlist=<machine .cfg file>
\end{lstlisting}
The ET will then be built using
\begin{lstlisting}
./simfactory/bin/sim build [<configurationname>] --thornlist=<thornlist>
\end{lstlisting}
The configuration name can be omitted, in which case it will default to ``sim''.
Different configurations, with different thornlists and configuration files, can co-exist.

To run, for example, the \verb|static_tov.par| parameter file and follow the
output, one would then execute

\begin{lstlisting}
./simfactory/bin/sim submit static_tov --configuration <configurationname> --parfile=par/static_tov.par --procs=1 --walltime=8:0:0
./simfactory/bin/sim show-output --follow static_tov
\end{lstlisting}
In this example we created a simulation called ``static\_tov'', only one
processor was requested and the simulation will abort after the specified
walltime of 8 hours.  As for building, if
\verb|--configuration <configuration>| is omitted, the default ``sim'' configuration is used.

This specific parameter file sets up a static TOV star (a model of a single
neutron star) with a mass of $1.4$ solar masses, and integrates the combined
relativistic fluid dynamics and spacetime evolution equations in time. The
spacetime is evolved using the BSSN 3+1 formulation of Einstein's equations and
the fluid is evolved using a high resolution shock capturing method.

To check the status of the simulation, use
\begin{lstlisting}
./simfactory/bin/sim list-simulations
\end{lstlisting}
If you accepted the default values at the setup stage, simulations will run in
the \verb|$HOME/simulations| folder (which needs to be created beforehand).

\subsubsection{Configuring additional machines}

Let us now see how to add an additional machine (say, your local workstation)
to SimFactory. The first step is adding the configuration
\verb|<machine name>.ini| file to \verb|./simfactory/mdb/machines/|. Such a file can be easily
adapted from the provided \verb|generic.ini|:
\begin{lstlisting}
[generic]

# Machine description
nickname        = generic
name            = Generic Machine
location        = somewhere
description     = Whatever
status          = personal

# Access to this machine
hostname        = generic.some.where
aliaspattern    = ^generic\.some\.where$

# Source tree management
sourcebasedir   = /home/@USER@
optionlist      = generic.cfg
submitscript    = generic.sub
runscript       = generic.run
make    	= make -j2
basedir         = /home/@USER@/simulations
ppn             = 1   # or more
max-num-threads = 1   # or more
num-threads     = 1   # or more
nodes           = 1
submit          = exec @SCRIPTFILE@ < /dev/null > /dev/null 2> /dev/null & echo $!
getstatus       = ps @JOB_ID@
stop            = kill @JOB_ID@
submitpattern   = (.*)
statuspattern   = "^ *@JOB_ID@ "
queuedpattern   = $^
runningpattern  = ^
holdingpattern  = $^
exechost        = echo localhost
exechostpattern = (.*)
stdout          = cat @SIMULATION_NAME@.out
stderr          = cat @SIMULATION_NAME@.err
stdout-follow   = tail -n 100 -f @SIMULATION_NAME@.out @SIMULATION_NAME@.err
\end{lstlisting}
For the most straightforward cases, it should be enough changing the Machine
description, hostname, aliaspattern and optionlist. Examples for optionlists
can be found in \verb|./simfactory/mdb/optionlists/|, and if none of these can
be used directly, a new list can be created and placed within that directory.

Once this is done, the newly added machine should appear in the list of known machines:
\begin{lstlisting}
./simfactory/bin/sim list-machines
\end{lstlisting}
and its configuration
\begin{lstlisting}
./simfactory/bin/sim print-mdb <machine>
\end{lstlisting}
We should now be ready to synchronize our files
\begin{lstlisting}
./simfactory/bin/sim sync <machine>
\end{lstlisting}
Once that is done, we repeat the building process from before, but with the \verb|--remote <machine>| argument:
\begin{lstlisting}
./simfactory/bin/sim --remote <machine> build [<configuration name>] --thornlist=<thornlist>
\end{lstlisting}

All SimFactory commands accept the \verb|--remote| option, taking a machine
name as argument. This allows one to build, submit and manage simulations on
several machines without ever having to directly login.

For more information, check~\cite{SimFactory:web}.

\section{Examples}
\label{sec:examples}
So far, we described core parts of the Einstein Toolkit as generally as possible
to give a consistent overview of its capabilities and components. Quite often
however, simple, additional examples can help to provide deeper understanding.
This is especially true in a workshop-like setting where teaching time is
usually scarce. Therefore, we present in this section three very hands-on
examples of how to use the Einstein Toolkit.

The first two examples, an evolution of a black hole binary, and the evolution
of a stable star, only present how users would set parameters to obtain the
respective simulations, using existing code and without the need to change it.
In contrast to these two, the third and last example show-cases an
implementation of the wave-equation, in particular using \codename{MoL}.

\subsection{Binary black hole coalescence}
\label{sec:bbh}
The ET provides, out-of-the-box, all the tools needed to evolve, with minimal input
from the user, a vacuum binary black hole coalescence.  We will here not
describe all the modules involved in such a simulation; we will merely briefly
go over the provided parameter file adapted for such an evolution and note the
parameters that a user would first change for slightly different physical
configurations.

Having obtained the ET as explained in the previous sections, a few example
parameter files are provided within the directory \verb|./Cactus/par/|.  In
particular, \verb|./Cactus/par/arXiv:1111.3344/bbh/BBHLowRes.par| (and corresponding higher
resolution versions) uses the thorn \codename{TwoPunctures}~\cite{Ansorg:2004ds}
to set up initial data for two black holes located at the $x$-axis with
(opposite) linear momentum along the $y$-axis.  This configuration is then
evolved using the \codename{ML\_BSSN} (\codename{McLachlan} BSSN) thorn.  \codename{AHFinderDirect}
searches for the black hole's apparent horizons at the designated intervals and
\codename{WeylScal4} extracts gravitational wave information. We should note
that modern initial data using the \codename{TwoPunctures} technique uses a few
more tricks to obtain more realistic initial data (e.g. by reducing the initial
eccentricity), but the simpler procedure presented here is sufficient to capture
the main ideas of such an evolution, and is therefore preferred in a workshop or
tutorial setting.

A streamlined, easier to follow, version of \verb|BBHLowRes.par| is provided at
\url{http://blackholes.ist.utl.pt/nrhep2/?page=material}, under the name
\verb|inspiral_d06_lres.par|. Let us now explore some of the parameters within
this file. We begin by noting the following group:
\begin{lstlisting}
ActiveThorns = "ReflectionSymmetry RotatingSymmetry180"
ReflectionSymmetry::reflection_z   = yes
CoordBase::domainsize = "minmax"
CoordBase::xmin =    0.00
CoordBase::ymin = -120.00
CoordBase::zmin =    0.00
CoordBase::xmax = +120.00
CoordBase::ymax = +120.00
CoordBase::zmax = +120.00
CoordBase::dx   =    2.00
CoordBase::dy   =    2.00
CoordBase::dz   =    2.00
\end{lstlisting}
This sets up the overall size of the numerical grid, which spans
$[[x_{\textrm{min}}, x_{\textrm{max}}], [y_{\textrm{min}}, y_{\textrm{max}}],
[z_{\textrm{min}}, z_{\textrm{max}}]] = [[0,120], [-120,120], [0,120]]$.
The problem has a mirror-symmetry in the plane of the inspiral, which is used
here to reduce the computational cost by a factor of $2$ by not evolving the
domain with $z<0$. In addition, the $\pi$-symmetry in the problem can be used
to only evolve points with $x>0$ (populating the missing part by rotating the
existing domain for $180$ degrees along the $z$-axis). This again reduces the
computational cost by about a factor of $2$, and is achieved by activating thorn
\codename{RotatingSymmetry180}.
The grid-spacing given here corresponds to the coarsest grid. The grid-spacing
of the inner regions (using AMR) will then depend on how many refinement levels
are chosen. Each level refines the region it is covering by a factor of two.
A user wanting to experiment with different grid sizes and/or
different resolutions would need to change these parameters accordingly.

The group
\begin{lstlisting}
CarpetRegrid2::num_centres = 2

CarpetRegrid2::num_levels_1         =  7
CarpetRegrid2::position_x_1         =  3.0
CarpetRegrid2::radius_1[ 1]         =  64.0
CarpetRegrid2::radius_1[ 2]         =  16.0
CarpetRegrid2::radius_1[ 3]         =   8.0
CarpetRegrid2::radius_1[ 4]         =   4.0
CarpetRegrid2::radius_1[ 5]         =   2.0
CarpetRegrid2::radius_1[ 6]         =   1.0
CarpetRegrid2::movement_threshold_1 =   0.16

CarpetRegrid2::num_levels_2         =  7
CarpetRegrid2::position_x_2         =  -3.0
CarpetRegrid2::radius_2[ 1]         =  64.0
CarpetRegrid2::radius_2[ 2]         =  16.0
CarpetRegrid2::radius_2[ 3]         =   8.0
CarpetRegrid2::radius_2[ 4]         =   4.0
CarpetRegrid2::radius_2[ 5]         =   2.0
CarpetRegrid2::radius_2[ 6]         =   1.0
CarpetRegrid2::movement_threshold_2 =   0.16
\end{lstlisting}
defines the number and radii of the inner refined regions, centered around each
of the black holes.  Should a user wish to change the initial position of the
black holes, it is important to also modify the
\verb|CarpetRegrid2::position_x_1|, \verb|CarpetRegrid2::position_x_2|
settings accordingly.

\begin{figure}
 \centering
 \includegraphics[width=0.3\textwidth]{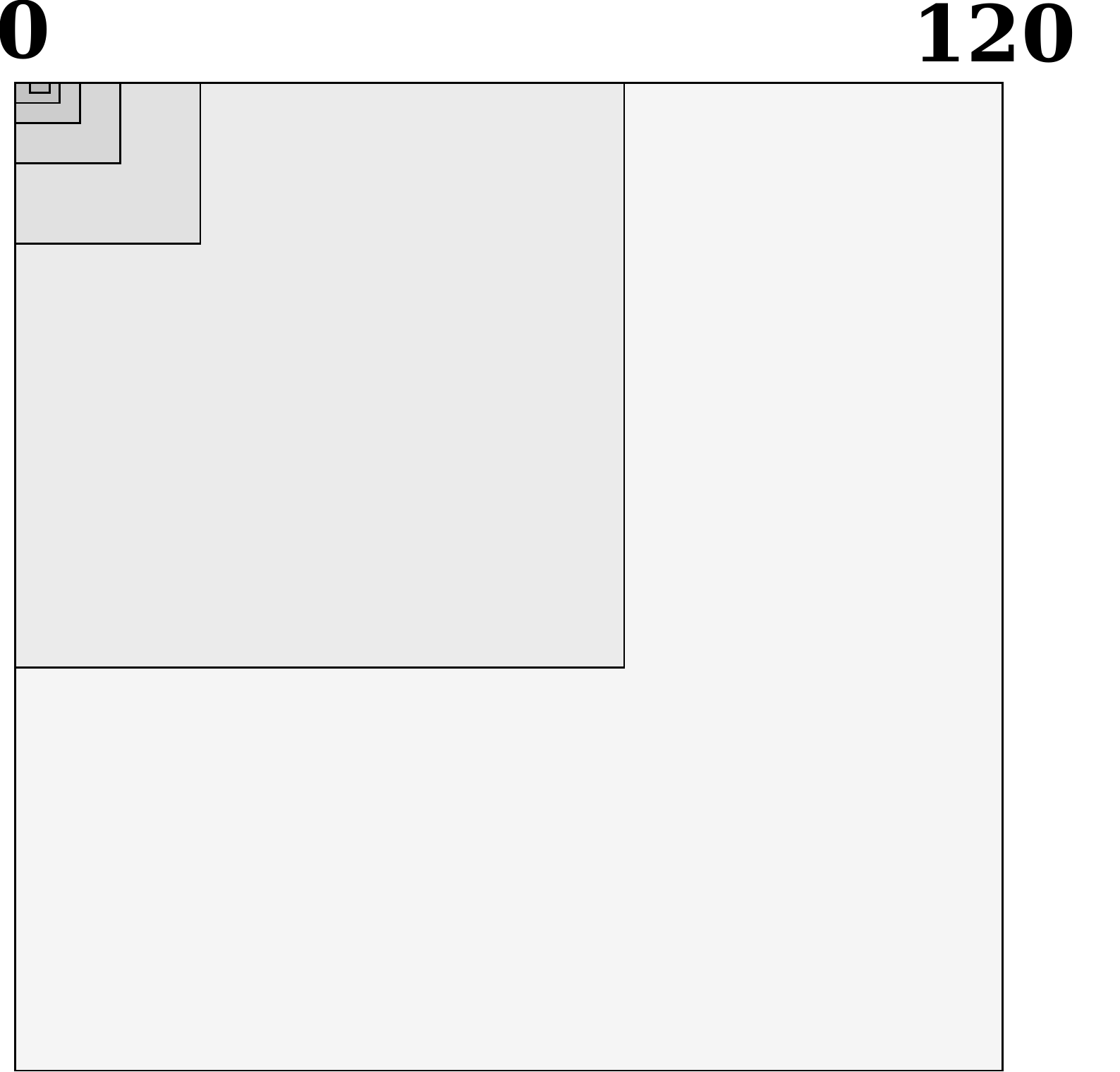}
 \includegraphics[width=0.3\textwidth]{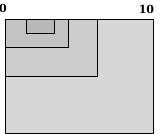}
 \caption{Sketch of the used grid setup. Note that only one of the finest
 region is shown because of the used symmetries. Also note that some of the
 overhead points are not shown, i.e. the ghost zones at the edge of refinement
 levels. On the right we show the same, but zoomed-in (with the outer-most three
 levels not shown). \label{fig:grid}}
\end{figure}
In Figure~\ref{fig:grid} we show the used grid setup in the $xy$ plane,
including all nested grids. Note that only one of the finest region is shown
because of the used symmetries.

The initial physical configuration is handled by the initial data thorn,
which in this case is \codename{TwoPunctures}.
The group
\begin{lstlisting}
TwoPunctures::par_b          =  3.001
TwoPunctures::par_m_plus     =  0.47656
TwoPunctures::par_m_minus    =  0.47656
TwoPunctures::par_P_plus [1] = +0.13808
TwoPunctures::par_P_minus[1] = -0.13808
\end{lstlisting}
defines the initial positions of the black holes to be at $(x,y,z) = (\pm
3.001,0,0)$ (by default, \codename{TwoPunctures} places the black holes
along the $x$-axis---use the parameter \codename{swap\_xz} to place them along the
$z$-axis), sets the black holes' ``bare mass'' parameter and its Bowen-York
linear momentum parameter.  This would be the first place to change should a
different physical configuration be needed.

When changing the initial position of the black holes, besides changing the
position of the inner refined regions, it is also important to correspondingly
change the initial guess for the apparent horizon finder:
\begin{lstlisting}
AHFinderDirect::origin_x                             [1] = +3.0
AHFinderDirect::initial_guess__coord_sphere__x_center[1] = +3.0

AHFinderDirect::origin_x                             [2] = -3.0
AHFinderDirect::initial_guess__coord_sphere__x_center[2] = -3.0
\end{lstlisting}

\begin{figure}
 \centering
 \includegraphics[width=0.9\textwidth]{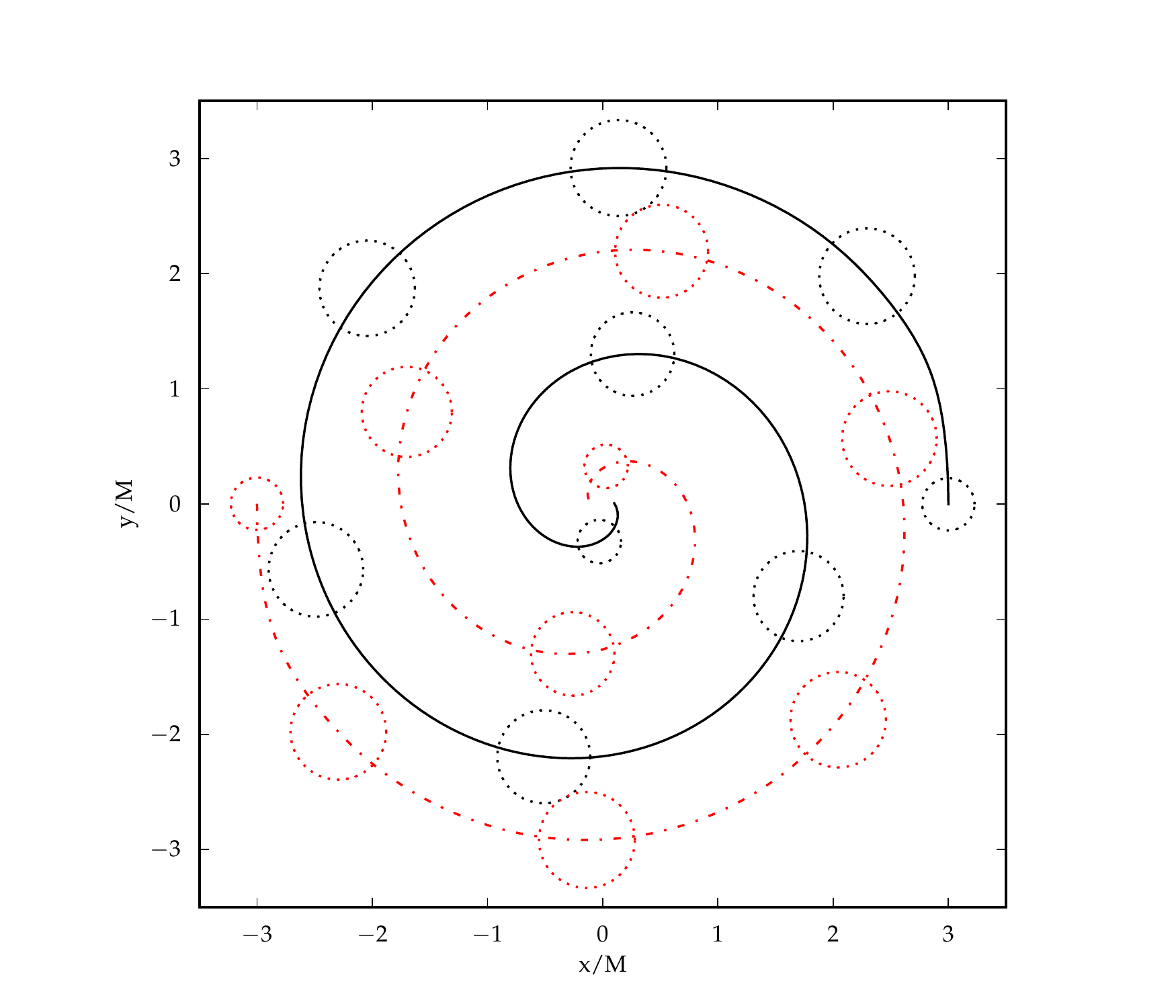}
 \caption{Puncture locations and mean radius of the
   corresponding apparent horizons (plotted every $15M$) during the binary
   inspiral. \label{fig:AH_traj}}
\end{figure}
In Figure~\ref{fig:AH_traj} we plot the evolution of the mean apparent horizon
radius of each black hole during the evolution of this setup. The script and
data files used to produce the plot are available at
\url{http://blackholes.ist.utl.pt/nrhep2/?page=material}.  An easy way to
produce a simpler version of the figure, also with \codename{matplotlib}, is the
following:
\begin{lstlisting}
ipython --pylab
  In [1]: xbh1, ybh1 = np.loadtxt("BH_diagnostics.ah1.gp", usecols=(2,3), unpack=True)
  In [2]: xbh2, ybh2 = np.loadtxt("BH_diagnostics.ah2.gp", usecols=(2,3), unpack=True)
  In [3]: axis("equal")
  In [4]: plot(xbh1, ybh1, linestyle="-", color="black")
  In [5]: plot(xbh2, ybh2, linestyle="-.", color="red")
\end{lstlisting}

\subsection{Simple TOV star}
\label{sec:tov}
The \codename{TOVSolver} routine in the Einstein Toolkit solves the standard
TOV\footnote{Tolman-Oppenheimer-Volkoff} equations \cite{Tolman:1939jz,Oppenheimer:1939ne} for the pressure,
enclosed gravitational mass $M_e$, and gravitational potential
$\Phi=\log\alpha$  in the interior of a spherically symmetric star in
hydro-static equilibrium. Details of the procedure, including equations,
can be found in~\cite{Loffler:2011ay}. We will use these initial data to
show linear oscillations of an otherwise stable TOV star.

The system setup by \codename{TOVSolver} is evolved using the BSSN evolution
system implemented in \codename{McLachlan} and the hydrodynamics evolution
system implemented in \codename{GRHydro}.

For the simple evolution presented here, we set up a stable TOV star described by a
polytropic equation of state with $K=100$ and $\Gamma=2$,
and an initial central density of $\rho_c=1.28\times10^{-3}$. This model can
be taken to represent a non-rotating NS with a mass of $M=1.4\mathrm{M}_\odot$,
and an EOS which mimics a cold neutron star reasonably well, at least to the
extend the EOS of such an object is currently known.

For reasonably accurate results of an evolution of such a system, runs are
typically performed with fixed mesh refinement, using about 5 levels of
refinement on a quadrant grid (symmetries provided by
\codename{ReflectionSymmetry} and \codename{RotatingSymmetry180}).  The outer
boundaries are placed at a few hundred ${M}$ (more then ten times the stellar
radius), and refined boxes are centered around the star at the origin, each
doubling the resolution.  Typical resolutions on the finest grid covering the
entire star are $0.500\mathrm{M}$ to $0.125\mathrm{M}$.

Evolutions of this size take much too long for a fast-paced tutorial
where participants are expected to run these on their own. Computational
resources might also be an issue if these resolutions would be tried. For such a tutorial setting, though, a much lower resolution and a smaller global grid
usually suffices, lowering the computational requirements and time
substantially.  It should, of course, be noted that results from such ultra-low
resolution runs are not to be trusted for scientific work, but that is not the
aim here. Increasing the parameters for resolution and grid size is an easy
task, best left to the interested participant, as follow-up after the workshop.
Thus, the evolution presented here only covers a global domain of $120\mathrm{M}$,
just about ten times the stellar radius. Only four levels of mesh refinement
are used, with a fine resolution of $1\mathrm{M}$. Note that this means that
there are only a few tens of points across the star diameter---probably one
of the lowest resolutions possible using the methods used here that still
produces somewhat sensible results.

\begin{figure}
 \centering
 \includegraphics[width=0.9\textwidth]{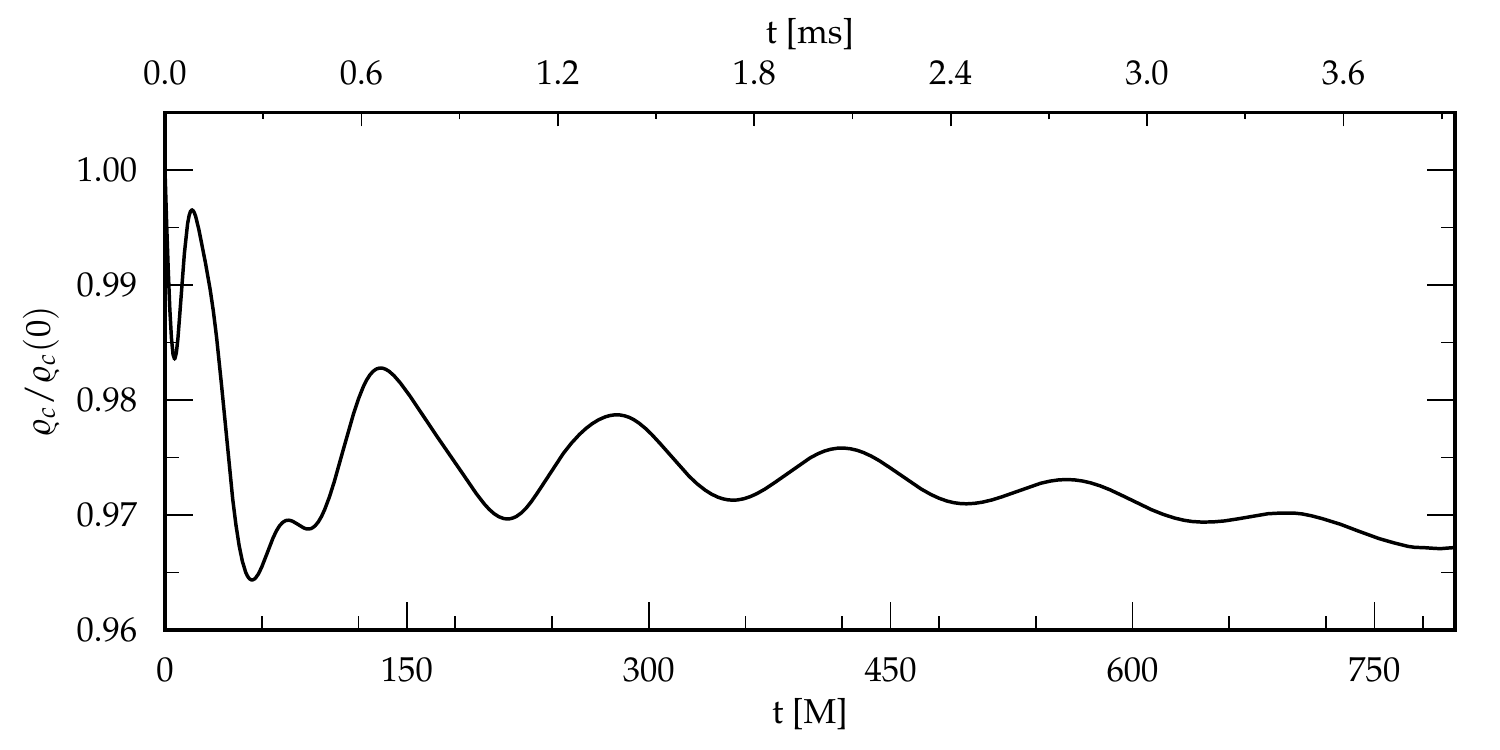}
 \caption{Evolution of the central density for the TOV star. Clearly visible is
   an initial spike, produced by the mapping of the one-dimensional equilibrium
   solution onto the three-dimensional evolution grid. Visible following this
   are damped oscillations of the central density of the
   star.\label{fig:tov_rho_max}}
\end{figure}

In Figure~\ref{fig:tov_rho_max} we show the evolution of the central density of
the star over an evolution time of $750\mathrm{M}$ ($3.6\mathrm{ms}$). The
initial spike is due to the perturbation of the solution resulting from the
mapping onto the evolution grid. After this (about $1\mathrm{ms}$), regular
star oscillations can be seen, damped mostly by adiabatic heating.

Figure~\ref{fig:tov_rho_max} represents a paper-quality plot of the central
density. During a workshop, quality requirements are much less stringent and
interactivity is much more important. One easy way to plot the same data
much quicker would be using \codename{matplotlib}:

\begin{lstlisting}
ipython --pylab
  In [1]: Fx, Fy = loadtxt("hydrobase::rho.maximum.asc", comments="#", usecols=(1,2), unpack=True)
  In [2]: plot(Fx, Fy/Fy[0])
\end{lstlisting}

As mentioned before, these results are in general to be considered quite poor.
However, the goal was not accuracy here. This becomes clear when considering the
computational requirements: about 400MB of memory and about 2.5 core-hours (2.5
hours on one current compute core). This code can take advantage of both multi-code
machines, as well as multi-node clusters (usually a combination of these), reducing
the overall physical runtime substantially.

\subsection{WaveMoL}
\label{sec:wavemol}
Let us explore a quite small thorn: \codename{CactusExamples/WaveMoL}.
While not a part of the Einstein Toolkit, this thorn illustrates the basic of
usage of the previously introduced \codename{MoL} (Method of Lines) \codename{Cactus}
thorn, which is used by \codename{McLachlan} and \codename{GRHydro}.

The \codename{MoL} thorn, already introduced in section~\ref{sec:mol}, provides
a convenient way to implement the Method of Lines technique inside \codename{Cactus}.  The
advantage of using this thorn is that one does not need to worry about writing
the time integration procedure itself; instead, one simply \emph{flags} the
variables that will be evolved and provide their ``right-hand-sides''.
\codename{MoL} will then take care of the evolution procedure using one of the
implemented methods, e.g., second, third of fourth order Runge-Kutta, or
Iterative Crank-Nicolson (to be chosen by the user at start-time).

For our example, let us then consider the wave equation 
\begin{equation*}
  \partial_t^2 \phi = \partial_{x^i}^2 \phi^i
\end{equation*}
To illustrate usage of the \codename{MoL} thorn, we rewrite the equations
in first order form, as was done in section~\ref{sec:mol}:
\begin{eqnarray}
  \partial_t \Phi & = & \partial_{x^i} \Pi^i, \label{eq:wave_eq1} \\
  \partial_t \Pi^j & = & \partial_{x^j} \Phi, \label{eq:wave_eq2} \\
  \partial_t \phi & = & \Phi, \label{eq:wave_eq3} \\
  \partial_{x^j} \phi & = & \Pi^j. \label{eq:wave_eq4}
\end{eqnarray}
The first three equations (five separate PDEs) will be evolved.  The final
equation is used to set the initial data and can be thought of as a constraint.
This determines how each of these variables has to be registered with
\codename{MoL}.

To obtain \codename{WaveMoL}, save the following text to a file
(called, e.g., \verb|WaveMol.th|)
\begin{lstlisting}
!CRL_VERSION = 1.0
!DEFINE ROOT = Cactus
!DEFINE ARR  = $ROOT/arrangements
!DEFINE COMPONENTLIST_TARGET = $ROOT/thornlists/
!DEFINE CACTUS_RELEASE = Cactus_4.1.0
!TARGET   = $ARR
!TYPE     = svn
!AUTH_URL = https://svn.cactuscode.org/arrangements/$1/$2/branches/$CACTUS_RELEASE
!URL      = http://svn.cactuscode.org/arrangements/$1/$2/branches/$CACTUS_RELEASE
!CHECKOUT =
CactusExamples/IDWaveMoL
CactusExamples/WaveMoL

\end{lstlisting}
and then use \verb|GetComponents| as previously:
\begin{lstlisting}
./GetComponents WaveMol.th
\end{lstlisting}

You should now have (under \verb|./Cactus/arrangements/CactusExamples/|) the
following directory structure
{\small\dirtree{%
.1 WaveMoL/.
.2 configuration.ccl.
.2 interface.ccl.
.2 schedule.ccl.
.2 param.ccl.
.2 src/.
.3 InitSymBound.c.
.3 Startup.c.
.3 WaveMoLRegister.c.
.3 WaveMoL.c.
.3 make.code.defn.
}}
The \verb|.ccl| files were already covered in section~\ref{sec:thorn_anatomy}, so we will here look into the C source code files.

We start with the \verb|WaveMoL_RegisterVars| routine, under \verb|WaveMoLRegister.c|:
\lstset{language=C}
\begin{lstlisting}
void WaveMoL_RegisterVars(CCTK_ARGUMENTS) {
  DECLARE_CCTK_ARGUMENTS;
  DECLARE_CCTK_PARAMETERS;
  CCTK_INT ierr = 0, group, rhs, var;

  group = CCTK_GroupIndex("wavemol::scalarevolvemol_scalar");
  rhs   = CCTK_GroupIndex("wavemol::scalarrhsmol_scalar");
  ierr += MoLRegisterEvolvedGroup(group, rhs);

  group = CCTK_GroupIndex("wavemol::scalarevolvemol_vector");
  rhs   = CCTK_GroupIndex("wavemol::scalarrhsmol_vector");
  ierr += MoLRegisterEvolvedGroup(group, rhs);

  var   = CCTK_VarIndex("wavemol::energy");
  ierr += MoLRegisterConstrained(var);
}
\end{lstlisting}
Note here how the variables
\verb|wavemol::scalarevolvemol_scalar| and
\verb|wavemol::scalarevolvemol_vector| are being flagged as \emph{evolved}
for \codename{MoL}, with right-hand side
\verb|wavemol::scalarrhsmol_scalar| and \verb|wavemol::scalarrhsmol_vector|
respectively. As we explained in section~\ref{sec:mol}, this allows the
\codename{MoL} thorn to evolve these grid functions from one time step to the
next and there is no need for the user to explicitly code any time-integration
routine.

For completeness, let us briefly look under the \verb|WaveMoL.c| file, where
we find the routine \verb|WaveMoL_CalcRHS|:
\lstset{language=C}
\begin{lstlisting}
void  WaveMoL_CalcRHS(CCTK_ARGUMENTS) {
  DECLARE_CCTK_ARGUMENTS;
  int i,j,k, index, istart, jstart, kstart, iend, jend, kend;
  CCTK_REAL dx,dy,dz, hdxi, hdyi, hdzi;

  /* Set up shorthands */
  dx = CCTK_DELTA_SPACE(0); dy = CCTK_DELTA_SPACE(1); dz = CCTK_DELTA_SPACE(2);
  hdxi = 0.5 / dx; hdyi = 0.5 / dy; hdzi = 0.5 / dz;
  istart = 1; jstart = 1; kstart = 1;
  iend = cctk_lsh[0]-1; jend = cctk_lsh[1]-1; kend = cctk_lsh[2]-1;

  /* Calculate the right hand sides. */
  for (k=0; k<cctk_lsh[2]; k++) {
    for (j=0; j<cctk_lsh[1]; j++) {
      for (i=0; i<cctk_lsh[0]; i++) {
        index = CCTK_GFINDEX3D(cctkGH,i,j,k);
        phirhs[index] = phit[index];
        phitrhs[index] = 0;
        phixrhs[index] = 0;
        phiyrhs[index] = 0;
        phizrhs[index] = 0;
      } 
    } 
  }  
  for (k=kstart; k<kend; k++) {
    for (j=jstart; j<jend; j++) {
      for (i=istart; i<iend; i++) {
        index = CCTK_GFINDEX3D(cctkGH,i,j,k);
        phitrhs[index] = 
            (phix[CCTK_GFINDEX3D(cctkGH, i+1, j, k)] -
             phix[CCTK_GFINDEX3D(cctkGH, i-1, j, k)]) *hdxi
          + (phiy[CCTK_GFINDEX3D(cctkGH, i, j+1, k)] -
             phiy[CCTK_GFINDEX3D(cctkGH, i, j-1, k)]) *hdyi
          + (phiz[CCTK_GFINDEX3D(cctkGH, i, j, k+1)] -
             phiz[CCTK_GFINDEX3D(cctkGH, i, j, k-1)]) *hdzi;
        phixrhs[index] = (phit[CCTK_GFINDEX3D(cctkGH, i+1, j, k)] -
                          phit[CCTK_GFINDEX3D(cctkGH, i-1, j, k)]) *hdxi;
        phiyrhs[index] = (phit[CCTK_GFINDEX3D(cctkGH, i, j+1, k)] -
                          phit[CCTK_GFINDEX3D(cctkGH, i, j-1, k)]) *hdyi;
        phizrhs[index] = (phit[CCTK_GFINDEX3D(cctkGH, i, j, k+1)] -
                          phit[CCTK_GFINDEX3D(cctkGH, i, j, k-1)]) *hdzi;
      } 
    }
  }
}
\end{lstlisting}
This routine computes the right-hand sides from equations~(\ref{eq:wave_eq1}--\ref{eq:wave_eq4})
using first order finite-differencing. These, as seen above, are then given to
the \texttt{MoL} thorn for the time-evolution.

Analogous, albeit more complicated procedures, are used within the
\codename{McLachlan} and \codename{GRHydro} thorns for evolving Einstein's field
equations and the dynamics of matter within.

\section{Final remarks}
\label{sec:remarks}

In these notes we hope to have provided a useful first guide for new users of
the Einstein Toolkit, an open-source computational infrastructure for numerical
relativity and relativistic astrophysics.  Its capabilities include:
\begin{itemize}
\item evolutions of vacuum spacetimes (e.g., binary black hole coalescence)
  through the BSSN general relativity spacetime evolution equations with standard
  ``moving puncture'' Gamma-driver and  1+log gauge conditions;
\item general-relativistic magneto-hydrodynamics evolutions, allowing, for
  example, simulations of magnetized isolated and binary neutron stars and
  collapsing stellar cores;
\item various initial data solvers or data importers for configurations of
     general relativistic objects;
\item tools for apparent and event horizon finding, black hole excision, gravitational
  wave extraction and spherical harmonics decomposition;
\item multidimensional Input/Output using HDF5, ASCII, Images;
\item Method of Lines time integration;
\end{itemize}

Tools and capabilities we have not explored include
\codename{Kranc}~\cite{Husa:2004ip,Kranc:web}---a Mathematica application that
converts continuum descriptions of PDEs into a \codename{Cactus} thorn; and
\codename{Llama}~\cite{Pollney:2009yz}---a 3-dimensional multiblock
infrastructure with adaptive mesh-refinement providing different patch systems
that cover the simulation domain by a set of overlapping patches.

We emphasize that we have here barely scratched the surface of the current
state of the toolkit and that large parts of it have intentionally been left
out. The goal of this work is not to provide an overview of all capabilities
of the ET, but to provide insight into the usage of the toolkit using a few
simple examples. Advanced users are then referred to more literature
covering parts of the toolkit, e.g.~\cite{Loffler:2011ay,Moesta:2013dna},
and are encouraged to seek direct contact to other active users of the
toolkit, e.g., using the ET mailing list \verb|users@einsteintoolkit.org|.

\section*{Acknowledgments}
We wish to thank the organizers and participants of the NR/HEP2 Spring School for the success of the event. We also thank Bruno Mundim, Joshua Faber and Yosef Zlochower for suggestions and Erik Schnetter and Manuela Campanelli for a careful reading of the manuscript.

M.Z. is supported by NSF grants AST-1028087, PHY-0969855, PHY-1229173, OCI-0832606.
The Einstein Toolkit is directly supported by the National Science Foundation
in the USA under the grant numbers 0903973 / 0903782 / 0904015 (CIGR\@) and
1212401 / 1212426 / 1212433 / 1212460 (Einstein Toolkit).  Computational
resources use regularly by the Einstein Toolkit maintainers for testing and
development includes resources provided by Louisiana State University
(allocations hpc\_cactus, hpc\_numrel and hpc\_hyrel), by the Louisiana Optical
Network Initiative (allocations loni\_cactus and loni\_numrel), by the National
Science Foundation through XSEDE resources (allocations TG-PHY060027N,
TG-ASC120003, TG-PHY100033, TG-MCA02N014, and TG-PHY120016).

Figures were generated with the \codename{Python}-based \codename{matplotlib}
package~\cite{Hunter:2007aa}.

\bibliographystyle{iopart-num-edit}
\bibliography{NRHEP2_ET,manifest/einsteintoolkit}

\end{document}